\documentclass{emulateapj}
\usepackage{color}
\usepackage{comment}
\usepackage{graphicx}

\usepackage{natbib,aas_macros}
\newcommand{\Lya}{Ly$\alpha$}
\newcommand{\OII}{[O\,{\sc ii}]}
\newcommand{\Ha}{H$\alpha$}
\newcommand{\Hb}{H$\beta$}

\newcommand{\OIII}{[O\,{\sc iii}]}

\newcommand{\fesc}{$f_{\rm esc}$}

\newcommand{\HI}{H{\sc i}}
\newcommand{\HII}{H{\sc ii}}

\newcommand{\ebv}{E(B$-$V)}
\newcommand{\qion}{$q_{\rm ion}$}

\newcommand{\xiion}{$\xi_{\rm ion}$}
\newcommand{\xiionzero}{$\xi_{\rm ion,0}$}
\newcommand{\nH}{$n_{\rm H}$}

\newcommand{\QHI}{$Q_{{\rm H}^{0}}$}

\slugcomment{Accepted for publication in ApJ Letters on Sep 22, 2016}

\shorttitle{
Hard ionizing spectrum in LAEs with intense \OIII\ emission
}
\shortauthors{Nakajima et al.}

\begin{document}

\title{
A Hard Ionizing Spectrum in $z=3-4$ Ly$\alpha$ Emitters with Intense [O\,{\sc iii}] Emission: \\
Analogs of Galaxies in the Reionization Era?
\altaffilmark{\dag}
\altaffilmark{\ddag}
}

\author{Kimihiko Nakajima\altaffilmark{1,2},
	Richard S. Ellis\altaffilmark{1,3}, 
	Ikuru Iwata\altaffilmark{4}, 
	Akio K. Inoue\altaffilmark{5}, 
	Haruka Kusakabe\altaffilmark{6}, 
	Masami Ouchi\altaffilmark{7,8}, and
	Brant E. Robertson\altaffilmark{9}	
      }

\email{knakajim@eso.org}

\altaffiltext{1}{%
European Southern Observatory,
Karl-Schwarzschild-Str. 2, 
85748, Garching bei Munchen, 
Germany
}
\altaffiltext{2}{%
Observatoire de Gen\`{e}ve, 
Universit\'{e} de Gen\`{e}ve, 
51 Ch. des Maillettes, 
1290 Versoix, 
Switzerland
}
\altaffiltext{3}{%
Department of Physics and Astronomy, 
University College London, 
Gower Street, London, 
WC1E 6BT, 
UK
}
\altaffiltext{4}{%
Subaru Telescope, 
National Astronomical Observatory of Japan, 
650 North A`oh\-{o}k\-{u} Place, 
Hilo, HI 96720, 
USA
}
\altaffiltext{5}{%
College of General Education, 
Osaka Sangyo University, 
3-1-1 Nakagaito, Daito, 
Osaka 574-8530, 
Japan
}
\altaffiltext{6}{%
Department of Astronomy, 
Graduate School of Science, 
The University of Tokyo, 
7-3-1 Hongo, Bunkyo-ku, 
Tokyo 113-0033, 
Japan
}
\altaffiltext{7}{%
Institute for Cosmic Ray Research, 
The University of Tokyo, 
5-1-5 Kashiwanoha, Kashiwa, 
Chiba 277-8582, 
Japan
}
\altaffiltext{8}{%
Kavli Institute for the Physics and Mathematics of the Universe (WPI), 
The University of Tokyo, 
5-1-5 Kashiwanoha, Kashiwa, 
Chiba 277-8583, 
Japan
}
\altaffiltext{9}{%
Department of Astronomy and Astrophysics, 
University of California, Santa Cruz,
1156 High Street
Santa Cruz, CA 95064,
USA
}

\altaffiltext{\dag}{%
Some of the data presented herein were obtained at the W.M. Keck Observatory, 
which is operated as a scientific partnership among the California Institute 
of Technology, the University of California and the National Aeronautics and 
Space Administration. The Observatory was made possible by the generous 
financial support of the W.M. Keck Foundation.
}

\altaffiltext{\ddag}{%
The observations were carried out within the framework of Subaru-Keck
time exchange program, where the travel expense was supported by 
the Subaru Telescope, which is operated by
the National Astronomical Observatory of Japan.
}

\begin{abstract}

We present Keck/MOSFIRE spectra of the diagnostic nebular emission lines 
\OIII$\lambda\lambda 5007,4959$, \OII$\lambda 3727$, and \Hb\ for a sample of 15 redshift
$z \simeq 3.1 - 3.7$ \Lya\ emitters (LAEs) and Lyman break galaxies (LBGs).
In conjunction with spectra from other surveys, we confirm earlier indications that LAEs 
have a much higher \OIII$/$\OII\ line ratio than is seen in similar redshift LBGs. By
comparing their distributions on a \OIII$/$\OII\  versus R23 diagram, we demonstrate
that this difference cannot arise solely because of their 
lower metallicities but most likely is due to a harder ionizing spectrum. Using
measures of \Hb\ and recombination theory, we demonstrate, for a subset of our 
LAEs, that \xiion - the number of Lyman continuum photons per UV luminosity -
is indeed $0.2-0.5$\,dex larger than for typical LBGs at similar redshifts. Using photoionization
models we estimate the effect this would have on both \OIII$/$\OII\  and R23 and conclude
such a hard spectrum can only partially explain such intense line emission. The additional possibility 
is that such a large \OIII$/$\OII\ ratio is in part due to density rather than ionization bound nebular
regions, which would imply a high escape fraction of ionizing photons.  We discuss
how further observations could confirm this possibility. Clearly LAEs with intense \OIII\ emission 
represent a promising analog of those $z>7$ sources with similarly strong lines which are thought
to be an important contributor to cosmic reionization. 

\end{abstract}


\section{INTRODUCTION} \label{sec:introduction}

To understand how cosmic reionization occurred during the redshift range $6<z<10$, it is necessary to identify the responsible sources.
Due to the steep decrease of the number density of quasars with redshift at $z>3$ \citep{fan2004}, the currently popular
viewpoint is that star-forming galaxies played the dominant role in delivering the necessary ionizing photons into 
the intergalactic medium (IGM; \citealt{robertson2015}).

The key question is whether the typical output of ionizing photons from $6<z<10$ galaxies is sufficient. 
This requires knowledge of  (i) the UV radiation emerging from their stellar populations, 
defined by \citet{robertson2013} in terms of \xiion, 
the number of Lyman continuum (LyC) photons per UV($1500$\,\AA) luminosity
and (ii) the fraction \fesc\ of such LyC photons that can escape scattering within the galaxy 
and its immediate vicinity. Neither of these important quantities is currently constrained for early galaxies
so this is the primary uncertainty in claims that reionization is driven primarily by star-forming galaxies. 

The present Letter is motivated by finding observational evidence that the ionizing spectrum is harder, and 
that \xiion\ and \fesc\ both increase with redshift, particularly for low mass, metal-poor systems 
characteristic of those that likely dominate reionization. Intense nebular emission, e.g., of \OIII$\lambda 5007$, 
appears to be more common in high redshift galaxies \citep{schenker2013,smit2014,smit2015}
and this has been interpreted as evidence for a harder ionizing spectrum.  Although such emission lines 
are not directly accessible with ground-based spectrographs beyond $z\sim 5$, IRAC photometry can 
still trace their presence via an excess inferred in the SEDs. The most intense \OIII\ emitters at $z>7$ 
located via IRAC photometry have confirmed \Lya\ emission \citep{zitrin2015,roberts-borsani2015}
which suggests they may have already created significant ionized bubbles for which a high value of 
\fesc\ is necessary \citep{stark2016}.

Since neither \xiion\ nor \fesc\ cannot directly be observed beyond $z\simeq 6$, we seek to find and study analogs 
of the reionization sources at $z\simeq 3$, the highest redshift where direct measures of these quantities are 
possible. 
The inter-dependence of strong \OIII\ emission and the leakage of LyC photons was first evaluated in the context 
of photoionization models by \citet{NO2014}. They suggested that large values of the emission line ratio 
\OIII$\lambda 5007$$/$\OII$\lambda3727$ may indicate a high value of \fesc.
Furthermore, the \OIII$/$\OII\ ratio is sensitive to the ionization parameter \qion, 
which is, 
in ionization equilibrium, 
related to the ionizing photon production rate \QHI\ and the gas density \nH, such that 
\begin{eqnarray}
q_{\rm ion}\propto (Q_{{\rm H}^{0}} \times n_{\rm H})^{1/3}.
	\label{eq:qion}	
\end{eqnarray}
A correlation between \OIII$/$\OII\ ratio and \xiion\ is thus suggested.
%
%
Recent supports for these conjectures are provided by 
significant LyC radiation detected in nearby intense \OIII\ emitters \citep{izotov2016a,izotov2016b,schaerer2016}.

In this Letter, we use the Keck near-infrared spectrograph MOSFIRE to compare the physical properties of a 
representative sample of intense \OIII-emitting Lyman alpha emitters (LAEs) at $z\simeq 3$ with equivalent 
data for Lyman break galaxies (LBGs). Our goal is to understand their \OIII$/$\OII\ line ratios in terms of a hard
ionizing spectrum. Ultimately we seek to verify that such sources may also have a high \fesc\ and thus
represent valuable analogs of the star-forming galaxies responsible for cosmic reionization.

\begin{figure}
  \centering
    \begin{tabular}{c}
      \begin{minipage}{1.00\hsize}
        \begin{center}
          \includegraphics[width=0.99\columnwidth]{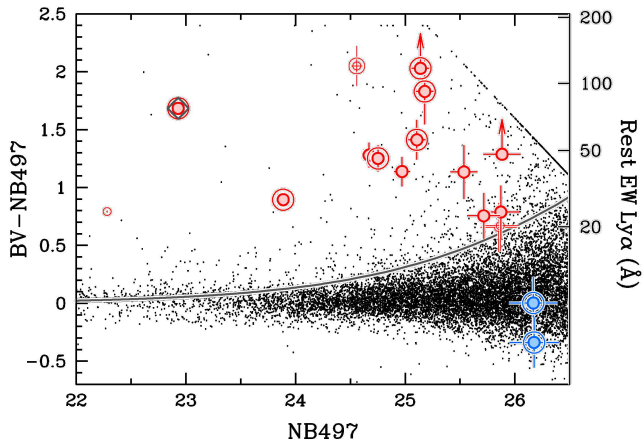}
        \end{center}
      \end{minipage}
    \end{tabular}
    \caption{%
  	Color-magnitude diagram of BV$-$NB497 vs. NB497 for the SSA22 field (where BV refers to ($2$B$+$V)$/3$). 
	Black dots show all the photometric detections. 
	The grey curve presents the $4\sigma$ photometric error in BV$-$NB497. 
	Large circles with error-bars show candidate $z\simeq 3.1$ LAEs (red), 
	LBGs (blue) and AGN-LAE (red with a grey diamond) observed with MOSFIRE.
	Filled and open circles present objects confirmed and unconfirmed respectively with MOSFIRE.
	Objects with prior optical spectroscopic confirmation are marked with a second circle.
	\label{fig:color_BVNB497_NB497}
    }
 \end{figure}

\section{Spectroscopic Data} \label{sec:data}

\subsection{Sample} \label{ssec:sample}

\begin{figure*}
  \centering
  \begin{tabular}{c}
    \begin{minipage}{0.47\hsize}
      \begin{flushright}
        \includegraphics[height=60mm]{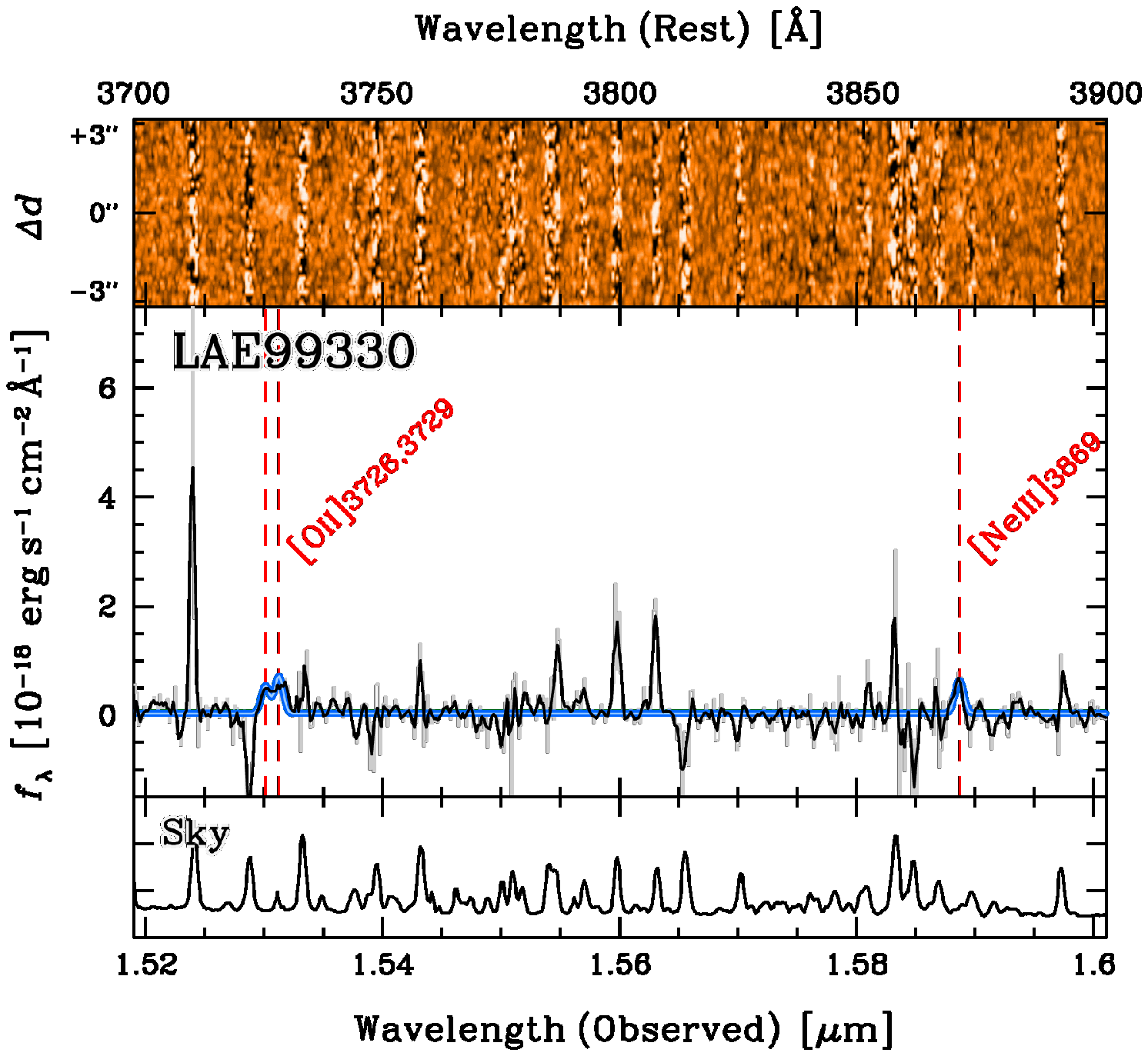}
      \end{flushright}
    \end{minipage} 
    \begin{minipage}{0.50\hsize}
      \begin{flushleft}
        \includegraphics[height=60mm]{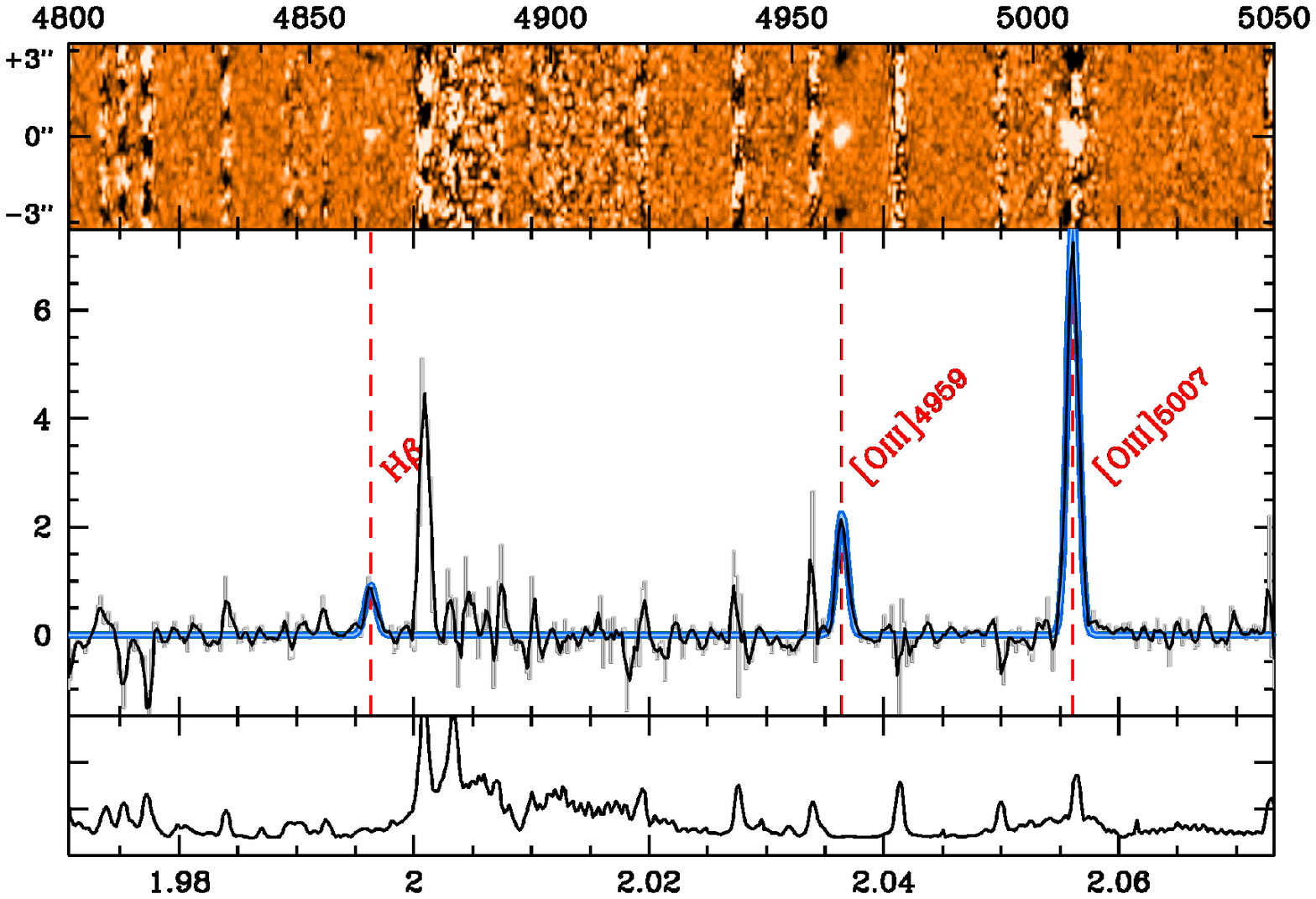}
      \end{flushleft}
    \end{minipage} 
   \end{tabular}
   \caption{%
    	MOSFIRE $H$ and $K$ spectra of 
	    a $z=3.105$ LAE (one of the brighter examples). 
        Detected emission lines are marked as vertical red-dashed lines,
	and the best-fit Gaussians are illustrated in blue.
	\label{fig:spec_MOSFIRE}
	}
 \end{figure*}

\begin{figure*}
  \centering
    \begin{tabular}{c}
      \begin{minipage}{1.00\hsize}
        \begin{center}
          \includegraphics[width=0.8\textwidth]{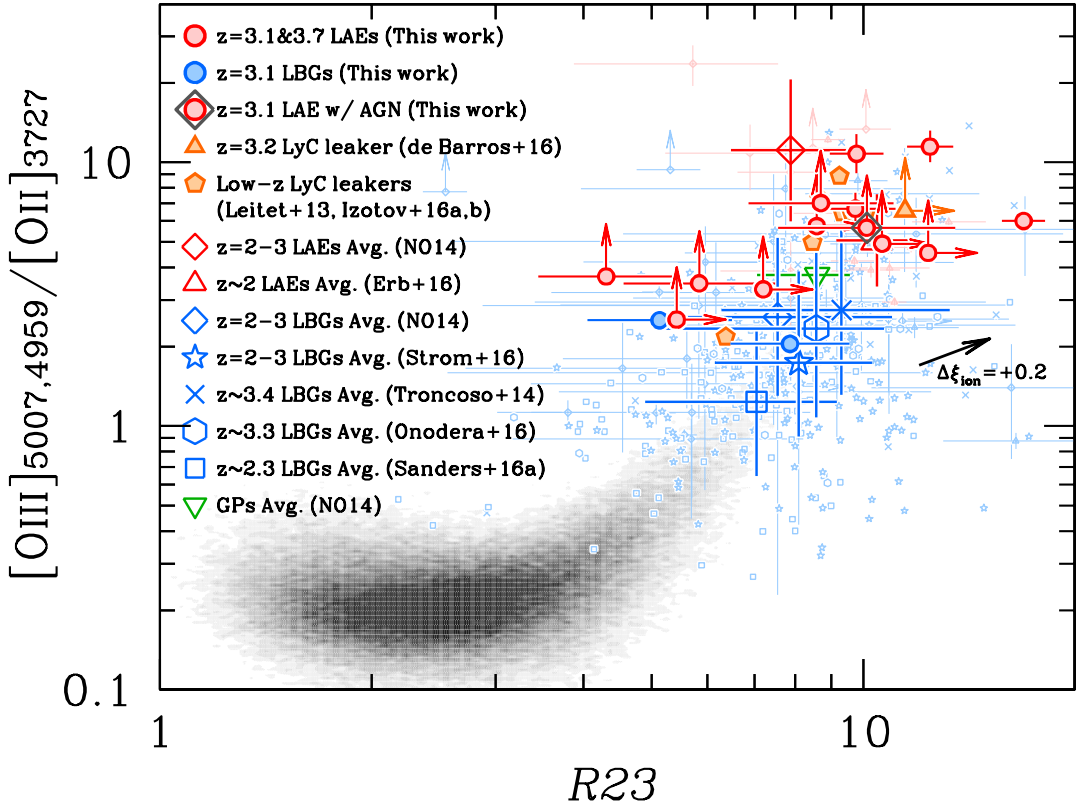}
        \end{center}
      \end{minipage}
    \end{tabular}
    \caption{%
  	\OIII$\lambda\lambda 5007,4959$$/$\OII$\lambda 3727$ line ratio versus R23-index for the MOSFIRE and other samples.
	Red and blue filled circles represent the newly-observed LAEs and LBGs, respectively;
	AGN86861 is shown with a grey-diamond. 
	Orange symbols show LyC leaking objects,
	and other red and blue symbols are high-$z$ LAEs and LBGs, respectively, compiled from the literature 
	as shown in the legend.
	Grey shading illustrates the equivalent distribution for SDSS galaxies. The black arrow indicates the shift expected for
	a harder ionizing radiation (see text).
	\label{fig:o3o2r23}
    }
 \end{figure*}

Our target sample is drawn from a Subaru imaging survey which has identified $z\simeq 3.1$ LAEs
in the SSA22 field (\citealt{hayashino2004,yamada2012,micheva2015}). LAEs at $z\simeq 3.1$ were selected
via their photometric excess in a narrow band filter centered at 497nm (Figure \ref{fig:color_BVNB497_NB497}). 
A limited amount of confirmatory optical spectroscopy has already
been conducted for this sample.
 
For the present campaign, the MOSFIRE pointing was chosen to include a few candidate LyC leakers 
on the basis of ground-based imaging. We then completed the sample of 16 targets including other 
LAEs representative of the parent sample.
We also included $2$ $z\simeq 3.1$ LBGs and one $z\sim 3.7$ LAE. 
The $z\sim 3.7$ LAE was initially classed as a LBG but optical spectroscopy revealed strong \Lya\ emission.
Among the total of 19 targets, $10$ were already confirmed from previous optical spectroscopy. 
One of the $z=3.1$ LAEs is a Type II AGN
\citep{micheva2016} 
but the other targets show no obvious indications of AGN activity.

\subsection{Observation and Data reduction} \label{ssec:observation}

Observations were undertaken on UT 2015 June 20 and 21. Both nights were photometric with a seeing 
of $0\farcs 45$--$0\farcs 50$. MOSFIRE multi-slit spectroscopy was taken using the $H$ and $K$ band filters
sampling the wavelength ranges of $1.45$--$1.78$ and $1.92$--$2.37$\,$\mu$m, respectively. Using a slit width 
of $0\farcs 7$ the resolving power is  $R\sim 3700$ in the $H$ band and $3600$ in $K$.  Individual exposures of 
$180$\,sec ($120$\,sec) were taken in $K$ ($H$) with a AB nod sequence of $3\farcs 0$ separation. The 
total integration time was $2.5$ hours in $H$ and $3.0$ hours in $K$.

Data reduction was performed using the MOSFIRE DRP%
\footnote{\url{https://keck-datareductionpipelines.github.io/MosfireDRP}. 
See also \citet{steidel2014}.}. The processing includes flat fielding, wavelength calibration, 
background subtraction and combining the nod positions. Wavelength solutions in $H$ were obtained 
from OH sky lines, while in $K$ a combination of OH lines and Ne arcs was used. 

Flux solutions and telluric absorption corrections were obtained from A0V {\it Hipparcos} stars  
observed at similar airmasses. 
This procedure also corrects for slit-losses since LAEs are unresolved in typical ground-based conditions 
\citep{malhotra2012}
and the standard stars were observed in a similar manner.
Flux calibrations were independently confirmed
using a relatively bright star ($K_{\rm Vega}=16.2$) placed on the mask. 

\subsection{Emission line identifications} \label{ssec:lines_identification}

One or more emission lines were detected at the $>3\sigma$ level in $17$ of our $19$ targets. 
Based on previously spectroscopic redshifts and the expected redshift of
$z\simeq3.1$ for the bulk of the sample, $15$ sources 
($12$ LAEs and two LBGs at $z\simeq 3.1$ and one LAE at $z\simeq 3.7$)
are readily identified with \OIII$\lambda 5007$. Two are considered to be \Ha\ at $z\sim 1.5-1.6$
and two further faint targets present no significant signal. The following analysis is therefore based on the
successfully confirmed $15$ sources. 

We measured \OII%
\footnote{We use the notation \OII$\lambda 3727$ as the sum of the doublet.}, 
\OIII\ and \Hb\  line fluxes by fitting a Gaussian profile to each line with the IRAF task \verb+specfit+ in stsdas.contrib.spfitpkg. 
In the fitting procedure, the redshift and FWHM of the \OIII$\lambda 5007$ (i.e., the strongest emission line) were 
adopted as Gaussians for the other lines. An H$+$K spectrum of a representative $z\simeq 3.1$ LAE with
a demonstrating of the fitting process is shown in Figure \ref{fig:spec_MOSFIRE}.

Table \ref{tbl:properties} lists the measured \OIII$/$\OII\ ratio and the R23-index,
(\OIII$\lambda\lambda 5007,4959$+\OII$\lambda 3727$)$/$\Hb,
for each of the $15$ targets.

\section{Results} \label{sec:results}

\begin{figure*}
  \centering
    \begin{tabular}{c}
      \begin{minipage}{1.00\hsize}
        \begin{center}
          \includegraphics[width=0.8\columnwidth]{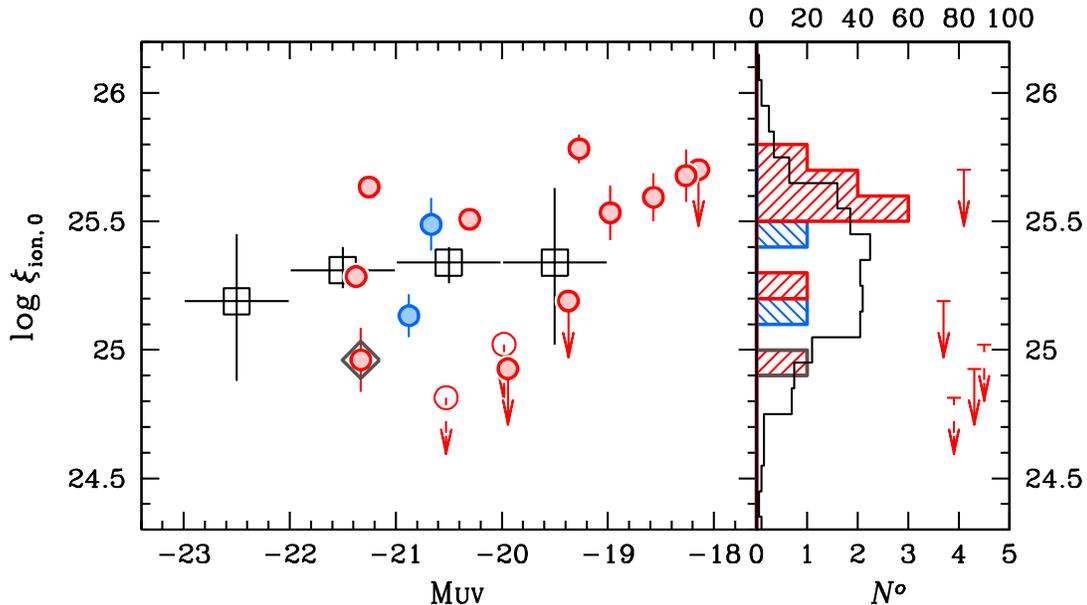}
        \end{center}
      \end{minipage}
    \end{tabular}
    \caption{%
	The ionization production parameter \xiionzero\ 
	as a function of UV absolute magnitude (left) 
	and its distribution (right; lower abscissa for our sample).
	Red and blue symbols refer to LAEs and LBGs respectively drawn
	from the current study. The grey-enclosed red symbol corresponds to AGN86861. 
	Red down-pointing arrows indicate $3\sigma$ upper-limits 
	for those sources for which \Hb\ is not detected 
	(open circles and dashed arrows show objects with less precise photometry). 
	The black points and histogram (upper abscissa) indicates the \xiionzero\ 
	for a larger sample of $z=3.8-5.0$ LBGs for which \Ha\ was inferred from 
	Spitzer photometry \citep{bouwens2015}.
	\label{fig:xi_MUV}
    }
 \end{figure*}

\subsection{The \OIII$/$\OII\ ratio of LAEs and LBGs} \label{ssec:o3o2}

Figure \ref{fig:o3o2r23} compares the results for our MOSFIRE $z\simeq$3.1 sample with other
high redshift and local sources in the \OIII$/$\OII\ ratio vs. the R23-index plane.
The latter index is a valuable probe of the gas-phase metallicity.
For the newly-observed sample, we assume no dust correction, an assumption we return to
in Section \ref{ssec:reddening}. In comparing with literature samples, we classify galaxies as LAEs 
if the rest EW(\Lya) exceeds $20$\,\AA\ (\citealt{NO2014,erb2016}), whereas for the 
continuum-selected galaxies 
at $z=2-4$ (\citealt{troncoso2014, sanders2016a,onodera2016,strom2016}), we assume these 
are dominated by LBGs  (cf. \citealt{shapley2003}). 
No obvious AGNs are included in the literature samples.

Since we are interested in the relevance of intense \OIII\ emission as a possible indicator
of a high \fesc\, we also plot recent galaxies revealing significant LyC emission; 
{\it Ion2} at $z=3.2$ \citep{debarros2016,vanzella2016}, five green pea galaxies at 
$z\simeq 0.3$ \citep{izotov2016a,izotov2016b}, and two local galaxies \citep{bergvall2006,leitet2013}.

Firstly, following \citet{NO2014}, it is clear that our LAEs have the highest \OIII$/$\OII\ ratio. 
A typical LBG at $z=2-4$ has an \OIII$/$\OII\ ratio of $\sim 1-3$, whereas the
LAEs presented here lie above $\simeq 3$ at the $>3\sigma$ level
with some as high as $\simeq 10$. Some LAEs may have even higher ratios given we cannot
always detect \OII\ . Notably, this LAE-LBG difference is seen even within our own sample.

Secondly, LAEs have an \OIII$/$\OII\ ratio higher than LBGs whose R23-indices are comparable. 
In local galaxies, a higher ionization parameter, \qion, is found in less chemically-enriched galaxies.
This correlation provides an empirical metallicity indicator for systems with strong line ratios 
(e.g., \citealt{maiolino2008}).
Although high-$z$ LBGs appear to follow the same relation (e.g., \citealt{shapley2015,sanders2016a,sanders2016b}),
LAEs lie above this trend indicative of an enhanced ionization parameter (e.g., \citealt{NO2014}). 
Galaxies have a metallicity of $\sim 0.2-0.3\,Z_{\odot}$ at a R23-index $\simeq 10$ \citep{maiolino2008} 
that is typically found in our LAE sample (see also \citealt{trainor2016}). 
On the other hand, if the \OIII$/$\OII\ ratio were adopted as a metallicity indicator \citep{maiolino2008}, 
the LAEs would have metallicities lower than those based on the R23-index and the \citet{KK2004} method 
by $0.4-1$\,dex.

\subsection{A Harder Ionizing Spectrum for LAEs} \label{ssec:xiion}

We now turn to estimating 
the ionizing photon production efficiency, 
\xiion, for our 
MOSFIRE sample using recombination lines in our spectra. This quantity 
represents the number of LyC photons per UV luminosity 
\citep{robertson2013, bouwens2015}; 
%
\begin{eqnarray}
& \xi_{\rm ion} &
	= \frac{Q_{{\rm H}^{0}}}{L_{\rm UV}} 
	\label{eq:xiion}	
\end{eqnarray}
%
where 
$L_{\rm UV}$ is the intrinsic UV-continuum luminosity.
%
The LyC photons production rate is determined by massive young stars
and best constrained by the hydrogen recombination lines,
in this case \Hb\.. We adopt \citet{bouwens2015} representation  
of the \citet{LH1995} relation
between \QHI\ and the \Hb\ luminosity $L$(\Hb), as 
%
\begin{eqnarray}
L({\rm H}\beta)\,\,[{\rm erg}\,{\rm s}^{-1}]
	= 1.36\times 10^{-12}\,Q_{{\rm H}^{0}} / 2.86 \,\,[{\rm s}^{-1}].
	\label{eq:LHbeta_NH0}	
\end{eqnarray}
%
%

Note that the conversion assumes no escaping ionizing photons, i.e
all are converted into recombination radiation. To recognize this, 
we adopt the notation \xiionzero, whose zero subscript indicates 
that the escape fraction of ionizing photons is zero.
The quantity \xiion\ can be derived by dividing \xiionzero\ by ($1-$ \fesc).
The UV luminosity is measured from an SED around $1500-1600$\,\AA\ using
the appropriate rest-frame broadband photometry assuming zero reddening.
The SED fitting was done as described in \citet{kusakabe2015}.

The assumption of zero reddening is important since it clearly
affects \xiion. We note \citet{schaerer2016} find dust correction 
decreases \xiion\ for LyC leakers at $z\simeq$0.3 by $\sim 0.3-0.4$\,dex.
\citet{castellano2014} report a similar trend for $z=3-4$ LBGs.
We return to this assumption in Section \ref{ssec:reddening}.

\Hb\ is robustly detected in $10$ of our $15$ galaxies ($7$ LAEs, $2$ LBGs and the AGN) 
and the derived values of \xiionzero\ are listed in Table \ref{tbl:properties}.
Among the $5$ LAEs whose \Hb\ is not detected, two have less precise photometry
and are removed from the following discussion as their \xiionzero\ values are less reliable. 
Figure \ref{fig:xi_MUV} compares the distribution of \xiionzero\ 
as a function of the absolute UV magnitude for the various
categories with those for $z=3.8-5.0$ LBGs analysed using a similar
approach based on inferred \Ha\ \citep{bouwens2015}.

Notwithstanding the 3 with upper limits, the \Hb-detected LAEs have an average \xiionzero\
larger than that inferred for $z=3.8-5.0$ LBGs by $\sim 0.2-0.5$\,dex.
They support the contention deduced from Figure \ref{fig:o3o2r23} that
LAEs typically have a harder ionizing radiation field than LBGs
(also refer to \citealt{matthee2016}).
The difference is apparent at a fixed UV magnitude suggesting a higher 
production rate of ionizing photons.

The offset could be even larger if, as suspected from the large \OIII$/$\OII\
ratio, \fesc\ is non-zero (see also \citealt{iwata2009,nestor2013,mostardi2013}).
For example, if \fesc\ $\simeq0.3$, \xiion\ becomes larger than \xiionzero\  by $\sim 0.15$\,dex. 
We note that two LAEs and one AGN-LAE have \xiionzero\ consistent with
the LBG population. This might arise from a particularly high \fesc, 
a less significant \Lya\ emission ($\simeq 20$\AA), and/or 
a more mature stellar population \citep{robertson2013} inferred from SED fitting ($>500$\,Myr).

\subsection{Constraints on Dust in LAEs} \label{ssec:reddening}

In the foregoing analyses, we have ignored any corrections for dust reddening in our sample which
would otherwise affect the results we present in both Figures \ref{fig:o3o2r23} and \ref{fig:xi_MUV}.

This assumption is supported by two LAEs in our sample, LAE93564 and LAE104037, for which
we are fortunate to see multiple Balmer lines and where the Balmer decrements 
are consistent with zero reddening, 
albeit with an uncertainty of $\Delta$\ebv\ $=0.1$.
%
For the other LAEs, SEDs based on our multicolor Subaru photometry further
restrict the range of \ebv\ values consistent with zero to within $1\sigma$. If we assume a median 
\ebv\ $=0.03$ consistent with the SEDs and assume an SMC attenuation law%
\footnote{As adopted in Figure \ref{fig:xi_MUV} for \citet{bouwens2015}'s \xiionzero\ measurements.}
\citep{gordon2003}, 
and the same color excess for the stellar and nebular emission, \xiionzero\ is only decreased by $\sim 0.1$\,dex. 
This would not change our conclusion that LAEs have a harder ionizing radiation field than LBGs.
Of course, any residual correction for dust reddening effect has an even smaller effect on the \OIII$/$\OII\ ratio 
(a decrease of $\sim 0.02$\,dex) and the R23-index (an increase of $<0.01$\,dex).

\section{Discussion} \label{sec:discussion}

We present Keck MOSFIRE measurements of the diagnostic nebular emission lines of \OIII, \OII, and \Hb\
drawn from a sample of $15$ $z\simeq 3.7$ and $3.1$ LAEs and LBGs.
In comparison with similar measurements of other surveys, we demonstrate 
that LAEs have much larger \OIII$/$\OII\ line ratios than those seen in LBGs.
This enhancement cannot be fully explained by their low metallicities, given the 
locally defined relation between metallicity and ionization parameter.
There are three possible explanations for this difference - 
(i) a larger gas density, 
(ii) a higher production rate of ionizing photons (and thus a larger mean \xiion), and 
(iii) a different geometry of \HII\ clouds (which relates to a higher \fesc).
For (i)-(ii),
following Equation (\ref{eq:qion}), a dense ISM or a 
high production of ionizing photons will increase the \OIII$/$\OII\ ratio.

We believe a denser ISM is an unlikely explanation because gas densities 
estimated from the resolved  \OII$\lambda 3729/\lambda 3726$ doublet line ratio
for LAEs are, on average, comparable to typical values for $z\sim 2.3$ 
continuum-selected galaxies
($\sim 200$\,cm$^{-3}$; \citealt{sanders2016a}). Although only two of our 
LAEs have a well-measured \OII\ doublet ratio, both indicate a modest gas density
of $50$--$300$\,cm$^{-3}$. 

A higher production rate of ionizing photons is a more natural explanation for 
the high \OIII$/$\OII\ ratio as we have directly verified they have a high \xiionzero.
However, the $0.3-0.5$\,dex larger \QHI\ observed would only increase
the ionization parameter $q_{ion}$ by $0.1-0.17$\,dex  and the \OIII$/$\OII\ ratio by $\sim 0.12-0.2$\,dex 
at a fixed metallicity of $\sim 0.2-0.3\,Z_{\odot}$ (i.e., R23 $\sim 10$; \citealt{KK2004}). 

Additionally,  the ionizing spectrum may be harder.
In this case we also expect an enhancement in the R23-index. We can calculate the magnitude 
of this effect using photoionization models similar to those presented in \citet{NO2014},
varying the stellar metallicity to represent a change in the hardness of the radiation
field (cf. \citealt{steidel2016}). We assume a gas phase metallicity of $Z\sim 0.5\,Z_{\odot}$ 
and an ionization parameter of log\,\qion\ $\sim 7.75$ \citep{sanders2016a,onodera2016}. 
To simulate the effect of a harder spectrum, we compare two stellar metallicity cases: 
$0.5Z_{\odot}$ comparable to the gas-phase metallicity (e.g., \citealt{kewley2013}) and a 
low value of $0.01Z_{\odot}$. With these assumptions, our photoionization models 
predict both the \OIII$/$\OII\ ratio and R23-index increase in the low metallicity case 
by only $0.1$\,dex, with \xiion\ enhanced by $0.2$\,dex. 
This change is illustrated in Figure \ref{fig:o3o2r23} with a black arrow.
This change of \xiion\ originates from the reduced UV-continuum since we fix the 
ionization parameter. If, conversely, the UV-continuum level is fixed,
the higher \xiion\ would further increase the ionization parameter by $\sim 0.06$\,dex
and \OIII$/$\OII\ ratio only by $\sim 0.08$\,dex. Neither is sufficient to explain
the high \OIII$/$\OII\ ratio. 
%

The third explanation would indicate a strong connection between
the large \OIII$/$\OII\ ratio and the escape fraction \fesc\ (\citealt{NO2014}). 
If \HII\ regions are density-bounded, the nebulae have a low column density 
of \HI\ and a high \fesc. The large \OIII$/$\OII\ ratio arises
since the outer zone that produces \OII\ is reduced while the inner \OIII-producing 
zone is unchanged (e.g., \citealt{kewley2013}). The R23-index is largely
unaffected since hydrogen will be ionized throughout the nebula.
For the ISM properties of a LBG, our photoionization models predict an \OIII$/$\OII\ ratio enhanced
by $0.4$\,dex with an unchanged R23-index ($0.02$\,dex decrease)
for \fesc\ of $30$\,\%.
This explanation is compatible with the modest densities observed in the LAEs,
since the \OII\ doublet probes the ionized gas density while density-bounded nebulae 
only affect the \HI\ density.

The only way to test whether such density-bounded \HII\ regions are 
dominant in high-$z$ LAEs and that this provides a contribution to the
large \OIII$/$\OII\ ratios, is to directly constrain the LyC leakage from $z\simeq$3 LAEs.
This would permit us to break the degeneracies in the discussion above. 
The prospects appear promising given the similarity in line emission
properties with the sources recently observed successfully with
LyC photons 
(\citealt{vanzella2016,izotov2016a,izotov2016b,schaerer2016}; 
see also \citealt{henry2015,verhamme2016}).
Regardless of the above degeneracy, our sample of $z\simeq3$ LAEs represent
valuable low redshift analogs of the $z>7$ sources with similarly intense 
\OIII\ emission and harder ionizing spectra
\citep{stark2015a,stark2015b,sobral2015,stark2016}
that may have the necessary high \fesc\ to drive cosmic reionization.

\begin{deluxetable*}{lccccccccc}
\tablecolumns{10}
\tabletypesize{\scriptsize}
\tablecaption{Spectral and stellar properties of the MOSFIRE-identified LAEs and LBGs%
\label{tbl:properties}}
\tablewidth{0.99\textwidth}
\tablehead{%
\colhead{Obj.} &
\colhead{EW(\Lya)} &
\colhead{spec(FUV)?} &
\colhead{$z_{\rm nebular}$} &
\colhead{\OIII$/$\OII} &
\colhead{R23} &
\colhead{log\,\xiionzero}  &
\colhead{log\,SFR$_{0}$} &
\colhead{log\,$M_{\star}$} &
\colhead{M$_{\rm UV}$} 
\\
\colhead{} &
\colhead{(\AA) \tiny (1)} &
\colhead{\tiny (2)} &
\colhead{\tiny (3)} &
\colhead{\tiny (4)} &
\colhead{\tiny (5)} &
\colhead{(Hz\,erg$^{-1}$) \tiny (6)} &
\colhead{($M_{\odot}$\,yr$^{-1}$) \tiny (7)} &
\colhead{($M_{\odot}$) \tiny (8)} &
\colhead{(AB) \tiny (9)} 
}
\startdata
LAE93564 & 
 $61^{+4}_{-4}$ &
 yes &
 $3.6768$ & 
 $10.78\pm 2.0$ & $9.8\pm 0.9$ &
 $25.63\pm 0.04$ &
 $1.80\pm 0.04$ &
 $10.28^{+0.27}_{-1.21}$ &
 $-21.3$
 \\
LAE94460$^{(\dag)}$ &  
 $54^{+9}_{-8}$ &
 yes &
 $3.0721$ & 
 $>2.5$ & $>5.4$ &
 $<25.0$ &
 $<0.68$ &
 $9.10^{+0.39}_{-0.73}$ &
 $-20.0$
 \\
LAE97081 & 
 $>213$ &
 yes &
 $3.0760$ &
 $>4.9$ & $>10.6$ &
 $<25.70$ &
 $<0.63$ &
 $8.08^{+0.70}_{-0.72}$ &
 $-18.1$ 
 \\
LAE97176 & 
 $62^{+15}_{-13}$ &
 yes &
 $3.0749$ &
 $>4.5$ & $>12.4$ &
 $<24.93$ &
 $<0.57$ &
 $9.28^{+0.31}_{-0.71}$ &
 $-19.9$
 \\
LAE103371 & 
 $151^{+72}_{-47}$ &
 yes &
 $3.0892$ &
 $>7.0$ & $8.7\pm 2.3$ & 
 $25.68\pm 0.10$ &
 $0.65\pm 0.10$ &
 $9.87^{+0.26}_{-0.72}$ &
 $-18.3$
 \\
LAE104037 &
 $37^{+3}_{-3}$ & 
 yes &
 $3.0646$ &
 $6.0\pm 0.3$ & $16.9\pm 1.2$ &
 $25.29\pm 0.03$ &
 $1.51\pm 0.03$ &
 $9.95^{+0.09}_{-0.13}$ &
 $-21.4$
 \\
LAE89723$^{(\dag)}$ & 
 $43^{+7}_{-7}$ &
 no &
 $3.1109$ &
 $>3.3$ & $>7.2$ &
 $<24.81$ &
 $<0.69$ &
 $8.47^{+1.06}_{-0.39}$ &
 $-20.5$
 \\
LAE91055 & 
 $>77$ &
 no &
 $3.0814$ &
 $>3.7$ & $4.3\pm 1.1$ &
 $25.59 \pm 0.09$ &
 $0.69\pm0.09$ &
 $8.80^{+0.82}_{-1.36}$ &
 $-18.6$
 \\
LAE97030 &
 $26^{+11}_{-9}$ &
 no &
 $3.0731$ &
 $6.7\pm 1.0$ & $9.7\pm 1.4$ &
 $25.78\pm 0.06$ &
 $1.16\pm 0.06$ &
 $8.41^{+0.68}_{-0.57}$ &
 $-19.3$
 \\
LAE97254 & 
 $74^{+27}_{-20}$ &
 no &
 $3.0709$ &
 $>3.5$ & $5.8\pm 1.6$ & 
 $25.53 \pm 0.11$ &
 $0.79\pm 0.11$ &
 $9.14^{+0.56}_{-0.41}$ &
 $-19.0$
 \\
LAE99330 & 
 $48^{+7}_{-6}$ &
 no &
 $3.1054$ &
 $11.5\pm 1.7$ & $12.4\pm 1.0$ &
 $25.51\pm 0.03$ &
 $1.30\pm 0.03$ &
 $10.64^{+0.33}_{-0.67}$ &
 $-20.3$
 \\
LAE104147 & 
 $24^{+7}_{-6}$ & 
 no &
 $3.0991$ &
 $>5.7$ & $>8.6$ & 
 $<25.19$ &
 $<0.61$ &
 $9.37^{+0.41}_{-0.83}$ &
 $-19.4$
 \\
AGN86861 &  
 $79^{+3}_{-3}$ &
 yes &
 $3.1051$ &
 $>5.6$ & $10.1\pm 3.4$ &
 $24.96\pm 0.12$ &
 $1.16\pm 0.13$ &
 $10.52^{+0.03}_{-0.04}$ &
 $-21.3$
 \\
LBG102826  & 
 $-4^{+4}_{-3}$ &
 yes &
 $3.0710$ & 
 $2.0\pm 0.2$ & $7.9\pm 1.7$ & 
 $25.13\pm 0.08$ &
 $1.15\pm 0.08$ &
 $9.80^{+0.08}_{-0.23}$ &
 $-20.9$
 \\
LBG104097  & 
 $-1^{+8}_{-6}$ &
 yes &
 $3.0671$ &
 $2.5\pm 0.2$ & $5.1\pm 1.4$ &
 $25.49\pm 0.10$ &
 $1.42\pm 0.10$ &
 $9.69^{+0.18}_{-0.16}$ &
 $-20.7$
\enddata
\tablecomments{
(1) Rest EW(\Lya).
For the $z\simeq 3.1$ objects, the EW is estimated from the BV$-$NB497 color. 
The EW of LAE93564 is derived from spectroscopy.
A $3\sigma$ lower-limit is given if the object is not detected significantly 
in the $BV$ image.
(2) Confirmed or not from previous rest FUV spectroscopy.
(3) Nebular redshift. 
(4) \OIII$\lambda\lambda5007,4959$$/$\OII$\lambda 3727$ ratio. 
(5) R23-index.
(6) 
\xiion\ under the assumption of a zero \fesc.
%
(7) SFR under the assumption of a zero \fesc, 
estimated from the \Hb\ luminosity.
An upper-limit of $3\sigma$ is adopted in Columns (4)--(7) 
if the line is not detected.
(8) Stellar mass derived from SED fitting,
which adopts an SMC dust law.
%
(9) Absolute UV magnitude measured from 
an SED around $1500-1600$\AA.
(\dag) The \xiionzero, stellar mass, and M$_{\rm UV}$ estimates are less certain 
due to less precise optical photometry. 
}
\end{deluxetable*}

\acknowledgments
We are grateful to the staff of the W. M. Keck Observatory who keep 
the instrument and telescope running effectively. 
We thank D. Stark, R. Bouwens, J. Matthee, A. Henry, R. Sanders, A.~E. Shapley, 
and the anonymous referee for useful comments.
The LAE catalog and some of the photometric and spectroscopic data
were kindly provided by T. Hayashino, T. Yamada, and Y. Matsuda,
and the \xiionzero\ measurements of $z=3.8-5.0$ LBGs by R. Bouwens.

KN acknowledges the JSPS Postdoctoral Fellowships for Research Abroad, 
and was benefited from a MERAC Funding and Travel Award.
RSE acknowledges support from the European Research Council through 
an Advanced Grant FP7/669253.

{\it Facilities:} %
\facility{Keck I (MOSFIRE)}




\begin{thebibliography}{}
\addcontentsline{toc}{chapter}{\bibname}
\expandafter\ifx\csname natexlab\endcsname\relax\def\natexlab#1{#1}\fi
\bibitem[\protect\citeauthoryear{Bergvall et al.}{2006}]{bergvall2006} Bergvall, N., et al.\ 2006, \aap, 448, 513
\bibitem[\protect\citeauthoryear{Bouwens et al.}{2015}]{bouwens2015} Bouwens, R.~J. et al.\ 2015, arXiv-eprints, arXiv:1511.08504
\bibitem[\protect\citeauthoryear{Castellano}{2014}]{castellano2014} Castellano, M., et al.\ 2014, \aap, 566, A19
\bibitem[\protect\citeauthoryear{de Barros et al.}{2016}]{debarros2016} de Barros, S., et al.\ 2016, \aap, 585, A51
\bibitem[\protect\citeauthoryear{Erb et al.}{2016}]{erb2016} Erb, D.~K., et al.\ 2016, arXiv-eprints, arXiv:1605.04919
\bibitem[\protect\citeauthoryear{Fan et al.}{2004}]{fan2004} Fan, X., et al.\ 2004, \apj, 128, 515
\bibitem[\protect\citeauthoryear{Gordon et al.}{2003}]{gordon2003} Gordon, K.~D., et al.\ 2003, \apj, 594, 279
\bibitem[\protect\citeauthoryear{Hayashino et al.}{2004}]{hayashino2004} Hayashino, T., et al.\ 2004, \aj, 128, 2073
\bibitem[\protect\citeauthoryear{Henry et al.}{2015}]{henry2015} Henry, A., et al.\ 2015, \apj, 809, 19
\bibitem[\protect\citeauthoryear{Iwata et al.}{2009}]{iwata2009} Iwata, I., et al.\ 2009, \apj, 692, 1287
\bibitem[\protect\citeauthoryear{Izotov et al.}{2016a}]{izotov2016a} Izotov, Y.~I., et al.\ 2016a, Nature, 529, 178
\bibitem[\protect\citeauthoryear{Izotov et al.}{2016b}]{izotov2016b} Izotov, Y.~I., et al.\ 2016b, \mnras, 461, 3683
\bibitem[\protect\citeauthoryear{Kewley et al.}{2013}]{kewley2013} Kewley, L.~J., et al.\ 2013, \apj, 774, 100
\bibitem[\protect\citeauthoryear{Kobulnicky \& Kewley}{2004}]{KK2004} Kobulnicky, H.~A., Kewley, L.~J.\ 2004, \apj, 617, 240
\bibitem[\protect\citeauthoryear{Kusakabe et al.}{2015}]{kusakabe2015} Kusakabe, H., Shimasaku, K., Nakajima, K., \& Ouchi, M.\ 2015, \apjl, 800, L29
\bibitem[\protect\citeauthoryear{Leitet et al.}{2013}]{leitet2013} Leitet, E., Bergvall, Hayes, M., Linn\'{e}, S., \& Zackrisson, E.\ 2013, \aap, 553, A106
\bibitem[\protect\citeauthoryear{Leitherer \& Heckman}{1995}]{LH1995} Leitherer, C., \& Heckman, T.~M.\ 1995, \apjs, 96, 9
\bibitem[\protect\citeauthoryear{Maiolino et al.}{2008}]{maiolino2008} Maiolino, R., et al.\ 2008, A\&A, 488, 463
\bibitem[\protect\citeauthoryear{Malhotra et al.}{2012}]{malhotra2012} Malhotra, S. et al.\ 2012, \apjl, 750, L36
\bibitem[\protect\citeauthoryear{Matthee et al.}{2016}]{matthee2016} Matthee, J., et al.\ 2016, arXiv-eprints, arXiv:1605.08782
\bibitem[\protect\citeauthoryear{Micheva et al.}{2015}]{micheva2015} Micheva, G., et al.\ 2015, arXiv-eprints, arXiv:1509.03996
\bibitem[\protect\citeauthoryear{Micheva et al.}{2016}]{micheva2016} Micheva, G., et al.\ 2016, arXiv-eprints, arXiv:1604.00102
\bibitem[\protect\citeauthoryear{Mostardi et al.}{2013}]{mostardi2013} Mostardi, R.~E., et al.\ 2013, \apj, 779, 65
\bibitem[\protect\citeauthoryear{Nakajima \& Ouchi}{2014}]{NO2014} Nakajima, K., \& Ouchi, M.\ 2014, \mnras, 442, 900
\bibitem[\protect\citeauthoryear{Nestor et al.}{2013}]{nestor2013} Nestor, D.~B., et al.\ 2013, \apj, 765, 47
\bibitem[\protect\citeauthoryear{Onodera et al.}{2016}]{onodera2016} Onodera, M. et al.\ 2016, \apj, 822, 42
\bibitem[\protect\citeauthoryear{Roberts-Borsani et al.}{2015}]{roberts-borsani2015} Robert-Borsani, G.~W., et al.\ 2015, arXiv-eprints, arXiv:1506:00854
\bibitem[\protect\citeauthoryear{Robertson et al.}{2013}]{robertson2013} Robertson, B., et al.\ 2013, \apj, 768, 71
\bibitem[\protect\citeauthoryear{Robertson et al.}{2015}]{robertson2015} Robertson, B., Ellis, R.~S., Furlanetto, S.~R., \& Dunlop, J.~S.\ 2015, \apjl, 802, L19
\bibitem[\protect\citeauthoryear{Sanders et al.}{2016a}]{sanders2016a} Sanders, R.~L., et al.\ 2016a, \apj, 816, 23
\bibitem[\protect\citeauthoryear{Sanders et al.}{2016b}]{sanders2016b} Sanders, R.~L., et al.\ 2016b, \apjl, 825, L23
\bibitem[\protect\citeauthoryear{Schaerer et al.}{2016}]{schaerer2016} Schaerer, D., et al.\ 2016, arXiv-eprints, arXiv:1606:00053
\bibitem[\protect\citeauthoryear{Schenker et al.}{2013}]{schenker2013} Schenker, M.~A., Ellis, R.~S., Konidaris, N.~P., \& Stark, D.~P.\ 2013, \apj, 777, 67
\bibitem[\protect\citeauthoryear{Shapley et al.}{2003}]{shapley2003} Shapley A.~E., et al.\ 2003, \apj, 588, 65
\bibitem[\protect\citeauthoryear{Shapley et al.}{2015}]{shapley2015} Shapley, A.~E., et al.\ 2015, \apj, 801, 88
\bibitem[\protect\citeauthoryear{Smit et al.}{2014}]{smit2014} Smit, R., et al.\ 2014, \apj, 784, 58
\bibitem[\protect\citeauthoryear{Smit et al.}{2015}]{smit2015} Smit, R., et al.\ 2015, \apj, 801, 122
\bibitem[\protect\citeauthoryear{Sobral et al.}{2015}]{sobral2015} Sobral, D., et al.\ 2015, \apj, 808, 139
\bibitem[\protect\citeauthoryear{Stark et al.}{2015a}]{stark2015a} Stark, D.~P., et al.\ 2015a, \mnras, 450, 1846
\bibitem[\protect\citeauthoryear{Stark et al.}{2015b}]{stark2015b} Stark, D.~P., et al.\ 2015b, \mnras, 454, 1393
\bibitem[\protect\citeauthoryear{Stark et al.}{2016}]{stark2016} Stark, D.~P., et al. 2016, arXiv-eprints, arXiv:1606.01304
\bibitem[\protect\citeauthoryear{Steidel et al.}{2014}]{steidel2014} Steidel, C.~C., et al.\ 2014, \apj, 795, 165
\bibitem[\protect\citeauthoryear{Steidel et al.}{2016}]{steidel2016} Steidel, C.~C., et al.\ 2016, \apj, 826, 159
\bibitem[\protect\citeauthoryear{Strom et al.}{2016}]{strom2016} Strom, A.~L., et al.\ 2016, arXiv-eprints, arXiv:1608.02587
\bibitem[\protect\citeauthoryear{Trainor et al.}{2016}]{trainor2016} Trainor, R.~F., et al.\ 2016, arXiv-eprints, arXiv:1608.07280
\bibitem[\protect\citeauthoryear{Troncoso et al.}{2014}]{troncoso2014} Troncoso, P., et al.\ 2014, \aap, 563, A58
\bibitem[\protect\citeauthoryear{Vanzella et al.}{2016}]{vanzella2016} Vanzella, E., et al.\ 2016, arXiv-eprints, arXiv:1602.00688
\bibitem[\protect\citeauthoryear{Verhamme et al.}{2016}]{verhamme2016} Verhamme, A., et al.\ 2016, arXiv-eprints, arXiv:1609.03477
\bibitem[\protect\citeauthoryear{Yamada et al.}{2012}]{yamada2012} Yamada, T., et al.\ 2012, \apj, 143, 79
\bibitem[\protect\citeauthoryear{Zitrin et al.}{2015}]{zitrin2015} Zitrin, A., et al.\ 2015, \apj, 810, L12
\end{thebibliography}
\end{document}